\documentclass[10pt,conference]{IEEEtran}
\usepackage{amsmath,amssymb}

\begin{document}

\title{Modified Euclidean Algorithms for Decoding Reed-Solomon Codes}

\author{
\IEEEauthorblockN{Dilip V. Sarwate}
\IEEEauthorblockA{Department of Electrical and Computer Engineering\\
and the Coordinated Science Laboratory\\
University of Illinois at Urbana-Champaign \\
Urbana, Illinois 61801 USA.\\
Email: {\tt sarwate@illinois.edu}}
\and
\IEEEauthorblockN{Zhiyuan Yan}
\IEEEauthorblockA{Department of Electrical and Computer Engineering\\
Lehigh University\\
 Bethlehem, Pennsylvania 18015 USA.\\
Email: {\tt yan@lehigh.edu}}}

\maketitle

\begin{abstract}
The extended Euclidean algorithm (EEA) for polynomial
greatest common divisors is commonly used in solving the
key equation in the decoding of Reed-Solomon (RS) codes,
and more generally in BCH decoding.  For this particular
application, the iterations in the EEA are stopped when the
degree of the remainder polynomial falls below a threshold.
While determining the degree of a polynomial is a
simple task for human beings, hardware implementation
of this stopping rule is more complicated.  This paper describes
a modified version of the EEA that is specifically adapted
to the RS decoding problem.  This modified algorithm requires
no degree computation or comparison to a threshold,
and it uses a fixed number of iterations.  Another advantage
of this modified version is in its application to the errors-and-erasures
decoding problem for RS codes where significant hardware
savings can be achieved via seamless computation.
\end{abstract}

\section{Introduction}
Reed-Solomon (RS) codes are among the
most widely used codes. Their applications range from consumer
electronics such as Compact Disc (CD) and Digital Versatile Disc
(DVD) players to optical communication and data storage systems.
Most high-throughput RS codec
architectures are based on reformulated versions
\cite{sarwateIEEE,zhang} of either the Berlekamp-Massey
algorithm \cite{berlekamp,blahut} or the extended
Euclidean algorithm (EEA)
\cite{berlekampchap,shao,shao88,sugiyama1,truong88,wilhelm}.
A variable number of iterations---at most $2t$ for correcting up
to $t$ errors---are used in most EEA-based decoders. One
exception is the errors-only hypersystolic Reed-Solomon decoder
proposed by Berlekamp \emph{et al.} \cite{berlekampchap} that uses
exactly $2t$ iterations regardless of the number of errors. Of
course, decoders based on the Berlekamp-Massey algorithm also
use exactly $2t$ iterations.

In this paper, new modifications of the EEA are proposed for both errors-only (EO)
decoding and errors-and-erasures (EE) decoding. First, a new modification
of the EEA for errors-only decoding is proposed based on the ideas in
\cite{berlekamp} and \cite{berlekampchap}. The new algorithm also uses
\emph{exactly} $2t$ iterations, thus eliminating the degree computation
and comparison in most variants of the EEA (see, for example,
\cite{baek}).   One particular hardware implementation
(not described in this paper) of this modified
algorithm turns out to be the \emph{same} circuit as that obtained by
implementing the RiBM algorithm of \cite{sarwateIEEE}, with the
difference that in one implementation the polynomials enter and
leave the circuit in ascending order of coefficients while in the other
implementation the polynomials enter and leave
in descending order of coefficients!  The RiBM algorithm is based
on the Berlekamp-Massey algorithm, and this result gives yet \emph{another}
equivalence between the Berlekamp-Massey algorithm and the EEA,
different from those that have been described previously
in \cite{dornstetter} and  \cite{heydtmann}.

The modified EEA proposed in this paper can be extended to
errors-and-erasures decoding, and used to
derive an errors-and-erasures decoding algorithm
that also iterates \emph{exactly} $2t$
times. This modified algorithm also combines the erasure-locator polynomial
computation and the solution of the modified key equation in a seamless
way. Hardware implementation of this algorithm eliminates not only the
degree computation and comparison (see, e.g., \cite{huangwu}) but also
the separate block used for computing the erasure-locator polynomial (see, for
example, \cite{zhang,KS97}), thus leading to considerable
savings.

\section{Notation and Preliminaries}
The codewords in a $t$-error-correcting cyclic Reed-Solomon
code \cite{blahut,berlekamp,wickerbhargava} of block length $n$
over $GF(2^m)$ are the polynomials
$C(z) = C_{n-1}z^{n-1} +  C_{n-2}z^{n-2} \ldots + C_1z + C_0$,
with the property that the $2t$ successive powers
$\alpha^{b_0}, \alpha^{b_0+1}, \ldots, \alpha^{b_0+2t-1}$
of $\alpha$, a primitive $n$-th root of
unity in $GF(2^m)$, are roots of $C(z)$.  Here, $b_0$ can
be any integer, but is often chosen to be $0$ or $1$
for ease of implementation.  The code has $n-2t$ information symbols.

\subsection{Errors-only Decoding Algorithms for Reed-Solomon Codes}
\label{section:erroronly}
Suppose the codeword polynomial $C(z)$ is transmitted and
the received word, corrupted by errors, is
$R(z) = C(z) + E(z)$
where $E(z)=\sum_{i=0}^{n-1} E_iz^i$.
The decoder computes the \emph{syndromes} of the error polynomial $E(z)$:
$S_j = R(\alpha^{b_0+j}) = C(\alpha^{b_0+j}) + E(\alpha^{b_0+j})
 = E(\alpha^{b_0+j})$, $0 \leq j < 2t.$
The \emph{syndrome polynomial} is defined as
$S(z) = S_0 + S_1z + \cdots + S_{2t-1}z^{2t-1}.$
If $\nu$ errors have occurred, the error polynomial $E(z)$ can
be written as
 $E(z) = Y_1z^{i_1} +  Y_2z^{i_2} + \cdots +  Y_\nu z^{i_\nu}$
where $Y_1, Y_2, \ldots, Y_\nu$, called the \emph{error values},
are said to have occurred at the \emph{error locations}
$X_1=\alpha^{i_1}, X_2=\alpha^{i_2}, \ldots, X_\nu=\alpha^{i_\nu}$
respectively.  The \emph{error-locator polynomial} $\Lambda(z)$ of
degree $\nu$ is defined to be
\begin{equation}
\Lambda(z) =  \prod_{j=1}^\nu (1 - X_jz) =
1 + \sum_{i=1}^{\nu} \Lambda_i z^i
\label{eq:errorlocator}
\end{equation}
while the \emph{error-evaluator polynomial} $\Omega(z)$ of degree
less than $\nu$ is defined as
\begin{equation}
\Omega(z)
 = \sum_{i = 1}^\nu Y_i {X_i}^{b_0}\prod_{j=1, j \ne i}^\nu (1 - X_jz)
= \sum_{i=0}^{\nu -1} \Omega_i z^i.
\label{eq:errorevaluator}
\end{equation}

\noindent The error-locator and error-evaluator polynomials defined above
are related to the syndrome polynomial by the \emph{key equation:}
\begin{equation}
\Lambda(z)S(z) \equiv \Omega(z) \bmod z^{2t}.
\label{eq:keyeq}
\end{equation}
Note that $S(z)$ is known to the decoder, while $\Lambda(z)$
and $\Omega(z)$ are not.  As the name suggests, solving the
key equation for both $\Lambda(z)$ and $\Omega(z)$ is the most
difficult part of the decoding process. In this paper, we focus
on the EEA algorithm \cite{blahut,sugiyama1} for solving the key equation.

After the key equation is solved, the errors can be corrected by
finding the error locations and computing the error values. The
error locations can be found via the \emph{Chien search}: for each $j$,
$0\le j \le n-1$, the decoder tests whether or not
$\Lambda(\alpha^{-j}) = 0$.  If $\Lambda(\alpha^{-j}) = 0$,
$E_j \neq 0$, that is, $j \in \{i_1, i_2, \cdots, i_\nu\}$.
The value of the $j$-th transmitted symbol is computed via
\emph{Forney's formula}:
\begin{equation}
C_j  = R_j + \frac{z^{b_0}\Omega(z)}{z\Lambda^\prime(z)}
\left |\!\begin{array}{l}\\{\scriptstyle{z = \alpha^{-j}}}\end{array}
\right .
\label{eq:forney}
\end{equation}
where
$\Lambda^\prime(z) =
\lambda_1 + 2 \lambda_2z + 3 \lambda_3z^2 +\cdots
 =  \lambda_1 + \lambda_3z^2 +\cdots$
is the formal derivative of $\Lambda(z)$.   It is worth
noting that most implementations of RS decoders compute
and use $\beta\Lambda(z)$ and $\beta\Omega(z)$ where $\beta$ is a
nonzero scalar whose value is immaterial:
$\beta\Lambda(z)$ has the same roots as $\Lambda(z)$ and
so the decoder finds the same error locations, and $\beta$
cancels out in (\ref{eq:forney}) and so the decoder finds the
same error values.  Henceforth, we ignore such scalar factors
in $\Lambda(z)$ and $\Omega(z)$.

\subsection{Errors-and-Erasures Decoding of Reed-Solomon Codes}
In some cases, the received words enter the decoder with some
symbols specially marked as being highly unreliable and hence
more likely to be in error than other symbols. These marked symbols
are called erasures. For a code with minimum Hamming distance
$d_{\text{\small{min}}}$, any pattern of $\mu$ erasures and
$\nu$ errors can be corrected \cite{blahut} as long as
$2\nu+\mu < d_{\text{\small{min}}}.$
Let $X_{1,e}, \cdots, X_{\nu,e}$ denote the (unknown) error
locations and $X_{1,\epsilon}, \cdots, X_{\mu,\epsilon}$ denote
the known \emph{erasure locations}. As before,
the error-locator polynomial is defined as
\[\Lambda _e(z)  =  \prod_{j=1}^\nu (1 - X_{j,e}z) \]
while the \emph{erasure-locator polynomial} is defined to be
\[\Lambda _\epsilon (z)  =  \prod_{j=1}^\mu (1 - X_{j,\epsilon}z) .\]
Note that $\Lambda _\epsilon (z)$ can be computed from the known
erasure locations whereas $\Lambda _e(z)$ is unknown. Similarly,
the error-evaluator polynomial is defined as
\[\Omega_e(z)  =
\sum_{i = 1}^\nu Y_{i,e}X_{i,e}^{b_0} \prod_{j=1, j \ne i}^\nu (1 - X_{j,e}z) \]
and the \emph{erasure-evaluator polynomial} is defined to be
\[\Omega_\epsilon (z)
=  \sum_{i = 1}^\mu Y_{i,\epsilon}X_{i,\epsilon}^{b_0}
\prod_{j=1, j \ne i}^\mu (1 - X_{j,\epsilon }z) \]
where $Y_{i,e}$ and $Y_{i,\epsilon}$ denote respectively the $i$-th
error and \emph{erasure values}. Also note that some
of the erasure values might be zero.

If we define the \emph{errata-locator polynomial} $\Lambda (z)$
of degree $\eta = \nu + \mu$
as $\Lambda _e(z)\Lambda _\epsilon (z)$
and the \emph{errata-evaluator polynomial} $\Omega (z)$
as $\Lambda _e(z)\Omega_\epsilon (z)+\Lambda _\epsilon (z) \Omega _e(z)$,
then the key equation (\ref{eq:keyeq}) still holds for the
errata-locator and errata-evaluator polynomials.  Furthermore, the $\eta$
errata locations can be obtained from the errata-locator
polynomial by the Chien search and the correct values of the
codeword symbols can be computed using Forney's formula.
Note that the errors-only decoding is simply the special case
of the errors-and-erasures decoding where
$\Lambda _\epsilon (z)  =1$ and $\Omega_\epsilon (z)  = 0$.
As in errors-only decoders, typical implementations compute
the same scalar multiple of all these polynomials, and the
value of this scalar does not affect the results
of any subsequent computations.

\subsection{Structure of RS Decoders}
As described above, the decoding of RS codes involve three
successive stages---syndrome computation (SC),
key equation solving (KES), and errata correction (EC). The
implementation of syndrome computation and errata correction,
which is described in, for example,
\cite{berlekamp}-\cite{blahut} is generally straightforward
and will not be discussed further in this paper. Here, we will focus on
the implementation of key equation solving, which is the most
difficult part of the decoding process.
\section{Modified EEA for Errors-only Decoding}
\label{section:eo}
\subsection{Key Equation Solution via the EEA}
\label{subsection:algI}
Sugiyama \emph{et al.} \cite{sugiyama1} first pointed out that
the extended Euclidean algorithm for computing the polynomial
greatest common divisor (GCD)  can
be used to solve
the key equation
(\ref{eq:keyeq}). The EEA, tailored
to solving the key equation, can be stated as follows:

\noindent{\bf EEA for Errors-Only Decoding: The EO Algorithm}
\begin{enumerate}
\item
\textbf{Initialization: Set} $v^{(0)} (z)\leftarrow z^{2t}$,
$v^{(1)} (z)\leftarrow S(z)$,\\
          $x^{(0)} (z) \leftarrow 0$, $x^{(1)} (z)\leftarrow 1$, and
$j\leftarrow 1$.
\item
\textbf{Iteration: While}
$\deg \left [ v^{(j)}(z) \right ] \geq t$, \\
$\text{~~~\bf Divide~} v^{(j-1)}(z)$ by $v^{(j)} (z)$ to obtain both\\
$\text{~~~~~the quotient~} q^{(j)}(z)\leftarrow \left \lfloor
{\frac{v^{(j-1)}(z)}{v^{(j)} (z)}} \right \rfloor$ and the \\
$\text{~~~~~remainder~}
v^{(j+1)}(z)\leftarrow v^{(j-1)}(z)-q^{(j)}(z)v^{(j)}(z)$.\\
$\text{~~~\bf Set~} x^{(j+1)}(z)\leftarrow x^{(j-1)}(z) - q^{(j)}(z)x^{(j)} (z)$.\\
$\text{~~~\bf Set~}j \leftarrow j+1$.

\item
      \textbf{Output: } $\Lambda (z)=x^{(j)} (z)$,
$\Omega (z)= v^{(j)}(z)$.

\end{enumerate}

\noindent
Let $k$ denote the value of $j$ when the EO algorithm
stops.  Then, the outputs $x^{(k)}(z)$ and $v^{(k)}(z)$ are
scalar multiples of $\Lambda (z)$ and $\Omega (z)$ as defined
in (\ref{eq:errorlocator}) and (\ref{eq:errorevaluator}) since
$x^{(k)}(0)$ is not necessarily $1$.   Also, it can be shown that the polynomials
$v^{(0)}(z), v^{(1)}(z), \ldots, v^{(k)}(z) = \Omega (z)$
computed by the EO algorithm have degrees $d_i$
that form a strictly decreasing sequence with
$d_0 = 2t$, $d_{k-1} = 2t - \nu,$  and $d_k = \nu$.

The drawbacks to efficient implementation of the above algorithm are
as follows.
\begin{itemize}
\item The degree $d_{j-1} - d_j$ of the quotient polynomial $q^{(j)}(z)$
can vary with $j$, and thus Step 2 of the EO algorithm
requires a variable number of
computations. This
complicates the control mechanism.  Furthermore, it is necessary to
divide the coefficients of $v^{(j-1)}(z)$ by the leading coefficient
of $v^{(j)} (z)$ in order to obtain the quotient polynomial $q^{(j)}(z)$.
\item Determining the stopping condition $\deg \left[v^{(j)}(z)\right] < t$
is difficult since data needs to be gathered
from many different cells in the circuit.
\end{itemize}
These two drawbacks have motivated many improvements.
\subsection{Partial Division and Cross-Multiplication}
Brent and Kung
\cite{brentkung} proposed a systolic array implementation of the polynomial GCD algorithm in which each of the polynomial division operations involved is
broken into a sequence of partial divisions, as humans often do in the
``long division'' method. In fact, this idea had been pointed out even earlier (see, for
example, \cite{berlekamp}).
Brent and Kung also proposed using cross
multiplications to avoid dividing one polynomial coefficient by another.
These notions can be explained as follows.
Let $U(z)$ and $V(z)$ denote polynomials of degrees $r$ and $s$
respectively where $r \geq s$.  Then, in the ``long division" of $U(z)$
by $V(z)$, the first step consists of subtracting
$\frac{U_r}{V_s}z^{r-s}V(z)$ from $U(z)$
to cancel out the highest degree term in $U(z)$.  If the remainder has
degree at least $s$, a different multiple of $V(z)$ is subtracted to cancel
out the highest degree term in the remainder, and so on.  But,
\begin{align}
\gcd(U(z), V(z)) &= \gcd(U(z) -\frac{U_r}{V_s}z^{r-s}V(z), V(z)) \label{eq:longdiv}\\
&= \gcd(V_sU(z) - U_rz^{r-s}V(z), V(z)) \label{eq:crossmult}
\end{align}
where (\ref{eq:crossmult}) follows from (\ref{eq:longdiv})
because $\gcd(A(z), B(z)) = \gcd(\beta A(z), B(z))$ for any nonzero
scalar $\beta$.  Thus, changing $U(z)$ to $V_sU(z) - U_rz^{r-s}V(z)$
instead of $U(z) - \frac{U_r}{V_s}z^{r-s}V(z)$ avoids a division while still
zeroing out the highest degree term in $U(z)$ and still having the
same GCD.  Since the computation of $x^{(j+1)}(z)$ in the EO algorithm is
of exactly the same form as the computation of $v^{(j+1)}(z)$,
a similar calculation can be used to update these polynomials as well.

These two basic ideas have been used in different ways by many
researchers to  design different algorithms for GCD computation
and RS decoding (see, for example,
\cite{berlekampchap,shao,shao88,truong88,wilhelm}).
All these algorithms actually compute scalar multiples $a \Lambda (z)$
and $a \Omega (z)$ of the error-locator and error-evaluator polynomials
defined in  (\ref{eq:errorlocator}) and (\ref{eq:errorevaluator}) respectively.
Our architectures also use the ideas of Brent and Kung, but compute
$a z^{i} \Lambda (z)$ and $a z^{i} \Omega (z)$ where $i \geq 0$.
Since the nonzero roots of $a z^i \Lambda (z)$ are the same as those
of $\Lambda (z)$ and the factors $a z^i$ cancel out in
Forney's formula (\ref{eq:forney}), such factors are
inconsequential and can be ignored.

As noted before \cite{berlekamp,berlekampchap,brentkung},
a polynomial division can be broken up into a sequence of partial divisions
for ease of implementation, and the cross-multiplication technique
can be used to avoid divisions of field elements
\cite{berlekampchap,brentkung,shao}.  The same ideas
can be adapted to eliminate the comparison
of $\deg [v^{(j)}(z)]$ with $t$ as well.  Our modification of
the EEA solves the key equation in \emph{exactly}
$2t$ steps; rather than in \emph{at most} $2t$ steps as in previous
work by others. When $\nu \leq t$ errors have occurred,
our algorithm computes $z^{\nu}\Lambda(z)$ and
$z^{\nu}\Omega (z)$ in $2\nu$ steps instead of $\Lambda(z)$
and $\Omega(z)$. Our algorithm
is also set up so that each of the additional $2t - 2\nu$ steps simply
multiplies the results by $z$ so that after a total
of $2t$ steps, our modification of the EEA has computed $z^{2t-\nu}\Lambda(z)$ and
$z^{2t-\nu} \Omega (z)$.  These give the same error locations
and error values as do $\Lambda(z)$ and $\Omega (z)$.
The advantages to our approach are that the degree checking is
avoided completely, and the key equation solution is produced with
a fixed latency, both of which properties simplify the control mechanism
in an implementation.

\subsection{The Modified EEA}
We claim that the following
modified version of the EEA solves the key equation
for RS decoding, producing polynomials
$X(z) = a z^{2t-\nu}\Lambda(z)$ and $V(z) = a z^{2t-\nu} \Omega (z)$.

\noindent{\bf Algorithm I (Modified Euclidean Algorithm)}
\begin{enumerate}
\item[\textbf{I.1}]
\textbf{Initialization:} $\delta \leftarrow 0$, $U(z)\leftarrow z^{2t}$,
$V(z)\leftarrow S(z)$, $W(z)\leftarrow 0$, and $X(z)\leftarrow 1$.

\item[\textbf{I.2}]
\textbf{Iteration: Repeat} $2t$ times:
\begin{enumerate}
\item \textbf{Set} $V(z) \leftarrow z V(z)$,  $X(z) \leftarrow z X(z)$,
 $\delta \leftarrow \delta -1$.
\item
       \textbf{If} $V_{2t} \ne 0$ and $\delta < 0$,\\
       $\text{\bf ~~~set~} \delta \leftarrow -\delta \text{~and~{\bf swap}~}
                            U \leftrightarrow V \text{~and~} W \leftrightarrow X.$
\item \textbf{Set}\vspace{-0.1in}
\begin{align*}
V(z) &\leftarrow  U_{2t} V(z) - V_{2t} U(z), \\
X(z) &\leftarrow  U_{2t} X(z) - V_{2t} W(z).
\end{align*}
\end{enumerate}

\item[\textbf{I.3}]
\textbf{Output:} $\Lambda(z)=X(z)$, $\Omega(z)=V(z)$, and $\delta$.
\end{enumerate}

\noindent If $\nu \leq t$ errors have occurred, then
after $2 \nu$ iterations of Step~I.2 in Algorithm~I,
$V(z) = z^{2t-d_{k-1}} v^{(j)}(z) = z^{\nu}\Omega(z)$ and
$X(z) = z^{2t-d_{k-1}} x^{(j)}(z) = z^{\nu}\Lambda(z)$
where scalar factors are ignored. When Step~I.2 is
iterated $2t - 2\nu$ more times, $X(z)$ and $V(z)$ are
multiplied by $z$ (Step 1.2a) and the ignorable scalar factor
 $U_{2t}$  (Step I.2c) each time.  Hence,
when Algorithm~I ends, $X(z) = z^{2t-\nu}\Lambda(z)$,
$V(z) =  z^{2t-\nu}\Omega(z)$, and $\delta = 2\nu - 2t \leq 0$.
If $\nu > t$, then Algorithm I terminates with $\delta > 0$.
In  practice, Steps~I.2a-I.2c are not executed in succession
but combined into a single calculation that computes
a Boolean control variable
$\text{SWAP} = (V_{2t-1} \neq 0) \wedge (\delta < 0)$ and then
\emph{simultaneously} sets
\begin{align*}
V(z) &\leftarrow U_{2t}zV(z) - V_{2t-1}U(z),\\
X(z) &\leftarrow U_{2t}zX(z) - V_{2t-1}W(z),\\
(U(z), W(z), \delta) &\leftarrow
\begin{cases}
(zV(z), zX(z), -\delta -1), &\text{if SWAP} = 1,\\
(U(z), W(z), \delta -1), & \text{if SWAP} = 0.
\end{cases}
\end{align*}
Note also that $\delta$ must be initialized to $-1$ for
this modified computation to work properly.  We refer
to this variation of Algorithm~I as Algorithm~I*.  The following
theorem summarizes the results of Algorithms~I and I*.\\
\emph{Theorem} 1:
If $\nu \leq t$ errors have occurred, then when Algorithm I or I* terminates,
$\delta = 2\nu - 2t - 1 < 0$ and
\begin{displaymath}
\begin{array}{llllllll}
& (X_{2t}, & X_{2t-1}, & \ldots, & X_{2t-\nu}, & X_{2t-\nu -1} & \ldots, & X_0)\\
= & (\beta\Lambda_{\nu}, & \beta\Lambda_{\nu -1}, & \ldots, & \beta\Lambda_0, & 0, &
\ldots, &  0),\\
& (V_{2t}, & V_{2t-1}, & \ldots, & V_{2t-\nu}, & V_{2t-\nu -1} & \ldots, & V_0)\\
 = & (0, & \beta\Omega_{\nu -1}, & \ldots, & \beta\Omega_0, & 0, &
\ldots, &  0).
\end{array}
\end{displaymath}
\noindent where $\beta$ is nonzero.  If Algorithm I or I* terminates with $\delta \geq 0$, then
more than $t$ errors have occurred and the error pattern
$E(z)$ is not correctable.
\section{Errors-and-Erasures Decoding}
\label{section:ee}
In errors-and-erasures decoding (see, for example,
\cite{sugiyama2,huangwu,blahut,zhang,KS97}), the key equation (\ref{eq:keyeq})
relating the errata-locator polynomial
$\Lambda(z) = \Lambda _e (z)\Lambda _\epsilon(z)$ and
the errata-evaluator polynomial $\Omega (z)$ is usually solved via the
following three steps executed in succession:
\begin{enumerate}
\item[1.] using the known erasure locations
  $X_{i,\epsilon}$, $1 \leq i \leq \mu$ to compute the
  erasure-locator polynomial $\Lambda _\epsilon(z)$ and
  the modified syndrome polynomial
  $$\hat{S}(z) \equiv \Lambda _\epsilon(z) S(z) \bmod z^{2t},$$
\item[2.] solving the modified key equation
  $$\Lambda _e(z) \hat{S}(z) \equiv \Omega (z) \bmod z^{2t}$$
  for the \emph{error}-locator polynomial $\Lambda _e (z)$ and
  the \emph{errata}-evaluator polynomial $\Omega (z)$
\item[3.] multiplying $\Lambda _e (z)$ by $\Lambda _\epsilon(z)$
  to obtain the errata-locator polynomial $\Lambda (z)$.
\end{enumerate}
The computations of $\Lambda _\epsilon(z)$ and $\hat{S}(z)$
can be implemented as $\mu$-iteration procedures
in which initial values $\Lambda _\epsilon(z) = 1$ and
$\hat{S}(z) = S(z)$ are multiplied successively by $(1 - X_{1,\epsilon}z)$,
$(1 - X_{2,\epsilon}z)$, \ldots, $(1 - X_{\mu,\epsilon}z)$.
Alternatively, $\Lambda _\epsilon(z)$ can be computed as described above
and then the polynomial product $\Lambda _\epsilon(z) S(z)$
computed in $\mu + 1$ further iterations (cf. \cite{zhang}).
Of course, if there are no erasures, then these calculations
do not need to be carried out.  Next, the (modified) key equation is
solved in \emph{at most} $2t-\mu$ iterations via a slightly
modified version of the extended Euclidean algorithm for
errors-only decoding. A slightly modified version of the Berlekamp-Massey
errors-only decoding algorithm also can be used for this purpose.
Finally, the last of the three steps above is not strictly
necessary, but is usually implemented (in fact, embedded
into the second step) because it is more convenient to use
$\Lambda (z)$ in computing errata values via Forney's formula.

It was pointed out by Blahut \cite{blahut} that if the registers
used to compute $\Lambda _e(z)$ are initialized to $\Lambda _\epsilon(z)$
instead of $1$, then the iterations during the solution of
the modified key equation produce
$\Lambda(z) = \Lambda _e (z)\Lambda _\epsilon(z)$ directly and thus
the third step above is in effect embedded into the
key equation solution.  Blahut \cite{blahut}
also noted that for the Berlekamp-Massey algorithm, it is unnecessary
to compute the modified syndrome polynomial: if the registers
used to compute $\Lambda _e(z)$ are initialized to
$\Lambda _\epsilon(z)$ instead of $1$, then the ``discrepancies'' calculated
in the Berlekamp-Massey algorithm are exactly those needed for
solving the modified key equation, and the algorithm produces
$\Lambda(z)$ directly instead of $\Lambda _\epsilon (z)$.
Unfortunately, reformulated Berlekamp-Massey algorithms such as
the riBM and RiBM algorithms of \cite{sarwateIEEE}
as well as \emph{all} key equation solvers that are based on the
extended Euclidean algorithm  \emph{do} need $\hat{S}(z)$.  However,
these algorithms are able to embed the third step above into
the key equation solution.  Finally, it has been noted by several
researchers that the operations used for the solution of the key
equation can be adapted to the computation of $\Lambda _\epsilon(z)$
or $\hat{S}(z)$ or both.  Thus,
the same hardware can be used in these calculations, which reduces
the number of finite-field multipliers required.
\subsection{Reformulation of Errors-and-Erasures Decoding Algorithms}
As pointed out in \cite{sugiyama2},
the modified key equation $\Lambda _e(z) \hat{S}(z) \equiv \Omega
(z) \bmod z^{2t}$  can be solved by using the extended Euclidean
algorithm shown below:

\noindent{\bf EEA for Errors-and-Erasures Decoding}
\begin{enumerate}
\item
 \textbf{Initialization: Set} $v^{(0)} (z)\leftarrow z^{2t}$, $v^{(1)} (z)\leftarrow \hat{S}(z)$,
          $x^{(0)} (z) \leftarrow 0$, $x^{(1)} (z)\leftarrow 1$, and $j\leftarrow 0$.
\item  \textbf{Iteration: While}
$\deg \left[v^{(j)}(z)\right] \geq t+\mu /2$,\\
$\text{~~~\bf Divide~} v^{(j-1)}(z)$ by $v^{(j)} (z)$ to obtain both\\
$\text{~~~~~the quotient~} q^{(j)}(z)\leftarrow \left \lfloor
{\frac{v^{(j-1)}(z)}{v^{(j)} (z)}} \right \rfloor$ and the \\
$\text{~~~~~remainder~}
v^{(j+1)}(z)\leftarrow v^{(j-1)}(z)-q^{(j)}(z)v^{(j)}(z)$.\\
$\text{~~~\bf Set~} x^{(j+1)}(z)\leftarrow x^{(j-1)}(z) - q^{(j)}(z)x^{(j)} (z)$.\\
$\text{~~~\bf Set~}j \leftarrow j+1$.

\item       \textbf{Output: } $\Lambda _e(z)=x^{(j)} (z)$,
$\Omega (z)= v^{(j)}(z)$.
\end{enumerate}

\noindent This algorithm is clearly similar to the EO algorithm
for errors-only decoding. In fact, the only differences between the two
algorithms are the initial values of $v^{(1)}(z)$ and the stopping condition.
Hence, direct implementation based on the above algorithm suffers the
same problems we described in Section III.A. Using the
same reformulation steps as in Section III, the above algorithm
can be modified to a $(2t-\mu)$-iteration algorithm that
eliminates the degree checking and produces $z^{c} \Lambda _e(z)$
and $z^{c} \Omega(z)$ instead of $\Lambda _e(z)$ and
$\Omega(z)$ respectively.

As mentioned above, the operations used to compute $\Lambda
(z)=\Lambda _e (z)\Lambda _\epsilon(z)$ in the third step can be
embedded in the modified Euclidean algorithm by initializing
$x^{(0)}(z)$ and $x^{(1)}(z)$ to scaled values $0\cdot \Lambda
_\epsilon(z)$ and $1\cdot \Lambda _\epsilon(z)$ respectively. Note
that the updates of $x^{(j)}(z)$ depend on $q^{(j)}(z)$, which are not
at all affected by the change in the initial values. Thus, each
$x^{(j)}(z)$ is scaled by $\Lambda _\epsilon(z)$, leading to a final
output $\Lambda _e (z)\Lambda _\epsilon(z)=\Lambda (z)$.

In order to combine the computation of $\Lambda _\epsilon(z)$ and
$\hat{S}(z)$ with the modified Euclidean algorithm (with the
computation of $\Lambda (z)$ embedded) into a single algorithm with
$2t$ iterations, we use a polynomial
$\psi (z)=\sum _{i=1}^{\mu}X_{i,\epsilon}z^{i-1} = \sum_{j=0}^{\mu-1} \psi_iz^j$
that can be formed easily during the syndrome
computation stage by saving the marked erasure locations.
We allow for the possibility that more than $2t$ erasures have
occurred, even though such an errata pattern is not decodable.
Our reformulated EEA for errors-and-erasures decoding is as follows:

\noindent{\bf Algorithm II}
\begin{enumerate}
\item
\textbf{Initialization:} $\delta \leftarrow -1$, $U(z)\leftarrow z^{2t}$,
$V(z)\leftarrow S(z)$,
  $X(z)\leftarrow 1$, $W(z)\leftarrow 0$,
and $\psi (z) =\sum _{i=1}^{\mu}X_{i,\epsilon}z^{i-1}$.
\item {\bf Iteration: Repeat} $2t$ times:\\
$\text{~~~{\bf Set} FIRST} \leftarrow (\psi_0\ne 0)$.\\
$\text{~~~SWAP}  \leftarrow (\neg \text{FIRST}) \wedge
(V_{2t-1}\ne 0) \wedge (\delta <0)$.\\
$~~~~(\gamma, \xi) \leftarrow \begin{cases} (U_{2t}, V_{2t-1}), &\text{~\,if FIRST } =0, \\
 (\psi_0, 1), &\text{~\,if FIRST } =1.
 \end{cases}$\\
$\text{~~~} \delta \leftarrow
\begin{cases}
-\delta - 1,  &\text{if SWAP } =1, \\
\delta - 1, &\text{if SWAP } = 0 \text{ and
  FIRST } = 0, \\
\delta, &\text{if FIRST } = 1.
\end{cases} $
\begin{align*}
\psi (z) &\leftarrow \displaystyle \biggr\lfloor \frac{\psi (z)}{z}\biggr\rfloor.  \\
V(z) &\leftarrow \gamma \cdot z V(z) -
 \begin{cases}
\xi \cdot U(z), &\text{~if FIRST } =0,\\
\xi \cdot V(z), &\text{~if FIRST } =1.
\end{cases} \\
X(z) &\leftarrow \gamma \cdot z X(z) -
  \begin{cases}
\xi \cdot W(z), &\text{if FIRST } = 0,\\
\xi \cdot X(z), &\text{if FIRST } = 1.
\end{cases}\\
U(z) &\leftarrow
 \begin{cases}
zV(z),  &\text{if SWAP } =1,\\
U(z), &\text{if SWAP } = 0.
\end{cases}\\
W(z) &= \begin{cases}
zX(z),  &\text{if SWAP } = 1,\\
W(z), &\text{if SWAP } = 0.
\end{cases}
\end{align*}
\vspace{-0.1in}
\item
\textbf{Output:} ${\Lambda}(z)=X(z)$, ${\Omega}(z)=V(z)$, $\delta$, and $\psi_0$.
\end{enumerate}

\noindent This algorithm uses a Boolean control variable FIRST that has value 1 only
when the erasure locations are being processed to compute
$\Lambda _\epsilon(z)$ and $\hat{S}(z)$. During this time, SWAP
is always $0$ and $\gamma$ is set to the erasure location being
processed currently. For each erasure location $\gamma$, Algorithm II sets $V(z)$
to $V(z)[1-\gamma z]$  and  $X(z)$
to $X(z)[1-\gamma z]$, thus obtaining
$\hat{S}(z)$ and $\Lambda _\epsilon(z)$ after all the $\mu$ erasure
locations have been processed one by one. Note that the update
$\psi(z) \leftarrow \bigr\lfloor \psi(z)/z\bigr\rfloor$ discards the
erasure location that was just processed and replaces it by the
next erasure location to be processed, and thus FIRST  becomes zero
after $\mu$ iterations. From this point onwards,
the updates of all the polynomials and of $\delta$ are the
same as in Algorithm I*,  and thus solve the modified
key equation in $2t-\mu$ iterations.
We remark that the received words with no erasures can be decoded
correctly by Algorithm II as well. In fact, Algorithm I*
corresponds to the special case of Algorithm II
where $\mu=0$ and FIRST is always $0$.
Similar to Theorem 1, we have\\
\emph{Theorem} 2:
Suppose that $\nu$ errors and $\mu$ erasures have
occurred where $2\nu+\mu \leq 2t$.  Let $\eta = \nu + \mu$
denote the total number of errata.  Then when Algorithm II terminates,
$\delta < 0$ and \\
\begin{displaymath}
\begin{array}{lllllllr}
& (X_{2t}, & X_{2t-1}, & \ldots, & X_{2t-\eta}, & X_{2t-\eta -1} & \ldots, & X_0)\\
= & (\beta\Lambda_{\eta}, & \beta\Lambda_{\eta -1}, & \ldots, & \beta\Lambda_0, & 0, &
\ldots, &  0),\\
& (V_{2t}, & V_{2t-1}, & \ldots, & V_{2t-\eta}, & V_{2t-\eta -1} & \ldots, & V_0)\\
 = & (0, & \beta\Omega_{\nu -1}, & \ldots, & \beta\Omega_0, & 0, &
\ldots, &  0).
\end{array}
\end{displaymath}
\noindent  where $\beta$ is nonzero.
If Algorithm II terminates with $\delta \geq 0$ or with
$\psi_0 \neq 0$, then
 the errata pattern is not correctable.

\section{Concluding Remarks}
In this paper, modified Euclidean algorithms that use \emph{fixed}
numbers of iterations are proposed for both errors-only and
errors-and-erasures decoding of RS codes. The salient feature of fixed
numbers of iterations leads to simpler control mechanisms and hence
hardware savings. The new algorithm for
errors-and-erasures decoding seamlessly combines the three steps
typically used in previously proposed architectures
into one procedure, leading to hardware savings.

\end{document}